\documentclass[aip, reprint]{revtex4-1}

\usepackage{graphicx}

\begin{document}

\title{Observation of neutron bursts produced by laboratory high-voltage atmospheric discharge} 

\author{A.V.~Agafonov}
\affiliation{P.N.~Lebedev Physical Institute of the Russian Academy of Sciences (FIAN), Moscow, 119991, Leninsky pr., 53}
\author{A.V.~Bagulya}
\affiliation{P.N.~Lebedev Physical Institute of the Russian Academy of Sciences (FIAN), Moscow, 119991, Leninsky pr., 53}
\author{O.D.~Dalkarov}
\affiliation{P.N.~Lebedev Physical Institute of the Russian Academy of Sciences (FIAN), Moscow, 119991, Leninsky pr., 53}
\affiliation{Centre for fundamental research (MIEM NRU HSE), Moscow, 101000, Myasnizkaya, 20}
\author{M.A.~Negodaev}
\affiliation{P.N.~Lebedev Physical Institute of the Russian Academy of Sciences (FIAN), Moscow, 119991, Leninsky pr., 53}
\author{A.V.~Oginov}
\email{oginov@lebedev.ru}
\affiliation{P.N.~Lebedev Physical Institute of the Russian Academy of Sciences (FIAN), Moscow, 119991, Leninsky pr., 53}
\author{A.S.~Rusetskiy}
\affiliation{P.N.~Lebedev Physical Institute of the Russian Academy of Sciences (FIAN), Moscow, 119991, Leninsky pr., 53}
\author{V.A.~Ryabov}
\affiliation{P.N.~Lebedev Physical Institute of the Russian Academy of Sciences (FIAN), Moscow, 119991, Leninsky pr., 53}
\author{K.V.~Shpakov}
\affiliation{P.N.~Lebedev Physical Institute of the Russian Academy of Sciences (FIAN), Moscow, 119991, Leninsky pr., 53}

\date{\today}

\begin{abstract}

Data on the observation of neutron bursts in the process of high-voltage discharge in the air at an average electric field strength $\sim 1$ MV$\cdot$m$^{-1}$ and discharge current $\sim 10$ kA are presented. Two independent methods (CR-39 track detectors and plastic scintillation detectors) registered neutrons within the range from thermal energies up to the energies above 10~MeV with the flux of $\gtrsim 10^6$~neutrons per shot into $4\pi$ solid angle.

\end{abstract}

\pacs{28.20.-v, 29.40.Gx, 52.80.Mg, 24.10.-i}

\maketitle

\section{Introduction}
Nowadays, the evidence of neutron flux enhancement in the lightning discharges in the atmosphere have been obtained in a number of experiments \cite{sha, gur, fle}. In the period of thunderstorm activity a significant excess of the neutron flux over the cosmic background is observed, and the registered neutron energy is in the range from the thermal (0.01~eV) to the fast one (tens of MeV). The nuclear fusion processes $^2$H($^2$H,n)$^3$He in the lightning channel \cite{kuz}, the reaction $^{12}$C($^2$H,n)$^{13}$N, $^{14}$N($^2$H,n)$^{15}$O \cite{fl2} and the photonuclear ($\gamma$,n) reactions have been considered \cite{bab} as possible mechanisms, leading to the formation of a statistically significant gain of a neutron flux in a thunderstorm atmosphere. However, a satisfactory explanation of the aggregate of the existing experimental results is still lacking. In this connection, the experiments on studying the neutron emission in the laboratory conditions, in high-voltage discharges in the air, which are similar to conditions observed in a natural storm, seem to be important.

\section{Experimental set-up}

\begin{figure}
\includegraphics{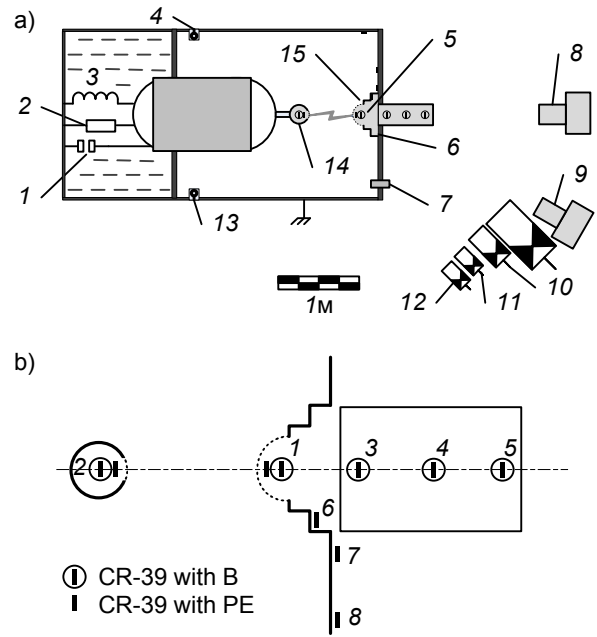}
\caption{ \label{fig:1}The scheme of laboratory experiment.  (a)
Layout of diagnostics: 1, 2 - capacitive and active dividers, 3 -
high voltage  input from Marx-generator, 4, 13 - magnetic probes,
5 - track detectors, 6 - anode shunt, 7 - Rogowski coil, 8,9 -
integral cameras, 10 - scintillation detectors, 11 - UV radiation
detector , 12 - PMT to visible light, 14 - cathode, 15 - anode.
(b) Layout of CR-39 track detectors: 1
- inside the anode, 2 - inside the cathode, 3, 4, 5 - axially
placed in water, 6, 7, 8 - radially placed at different distances
from the discharge.}
\end{figure}

The experiments have been carried out with a high current electron beam accelerator ERG, reconstructed for studying the high-voltage discharge in the air \cite{aga}. The scheme of the experiment is shown in Fig.~\ref{fig:1}a. The voltage amplitude up to 1~MW from the Marx-generator with the store energy of 60~kJ is fed through the entranced oil-air isolator at to duralumin flange with replaceable electrodes (cathode), and provides discharge current of $10 - 15$~kA in the air. The diameter of the external grounded cylindrical conductor limits the allowable distance between the cathode and the anode at the level of 1~m. The anode flange can be moved along the axis of the system for a smooth change of the gap. The anodes and cathodes used the same set of replaceable electrodes.

Electrophysical diagnostics included the recording of currents waveforms (anode shunt, Rogowski coils), the voltage (active and capacitive divider), and azimuthal magnetic field probes with a bandwidth of 200~MHz. The channel formation of the discharge is monitored using integral shooting in the optical range from two angles.

To register the integral neutron flux the track detectors CR-39, produced by Fukuvi Chemical Industry company, which are insensitive to the electromagnetic radiation, have been used. The track detectors CR-39 are located near the zone of discharge as shown in Fig.~\ref{fig:1}b.

The calibration of the CR-39 detector by charged particles has been carried out on the beam of protons from an electrostatic accelerator (E$_{\rm p} = 0.5 - 3.0$~MeV), with standard $\alpha$-sources (E$_\alpha = 2 - 7.7$~MeV), and on the beam of a cyclotron (E$_\alpha = 8 - 30$~MeV) at SINP MSU. After the irradiation the detectors were etched in the solution 6M NaOH in H$_2$O at the temperature of 70C during 7 hours. A detailed procedure of the track detector calibration is considered in \cite{bel}.

The neutron calibration of the CR-39 detector was performed using a $^{252}$Cf source with the activity of  $3\times10^4$ n$\cdot$s$^{-1}$ in the solid angle of 4$\pi$. A detector with the radiator of 120~$\mu$m polyethylene (detectors PE in Fig.~\ref{fig:1}) was used. Recoil protons produced by fast neutrons are registered by track detector. The calibration measurements have shown that the diameters of the proton tracks are in the range of $4 - 8$~$\mu$m. The average efficiency of fast neutron registration by CR-39 detector was $\eta_{\rm{n1}} = 6\times10^{-5}$.

For the registration of thermal neutrons by the reaction

\begin{equation}
^{10}\rm{B} + \rm{n} \longrightarrow ^7\rm{Li \:(0.8 \:MeV)} + ^4\rm{He \:(2 \:MeV)}
\label{eq1}
\end{equation}
the detectors were placed into the 20$\%$ solution of Na$_2$B$_4$O$_7$ in glycerol (detectors with B in Fig.~\ref{fig:1}). The neutrons were registered by counting the tracks of $\alpha$-particles with E$_\alpha <$ 2 MeV, which, according to the calibration, have the diameters of $10 - 12$~$\mu$m. The average efficiency of thermal neutron registration by CR-39 detector was $\eta_{\rm{n2}} = 1.4\times10^{-6}$.

Fast neutrons of the energy E$_{\rm{n}} > 10$~MeV was detected by the reaction of

\begin{equation}
^{12}\rm{C + n} \longrightarrow 3\alpha + n',
\label{eq2}
\end{equation}
with the energy threshold of about 10~MeV. A characteristic signature of the desintegration of the nucleus $^{12}$C presents three $\alpha$-particles, and their tracks come from a single point. The average efficiency of registration of the fast neutrons by reaction (\ref{eq2}) using the CR-39 detector with a 120~$\mu$m PE radiator was $\eta_{\rm{n3}} = 1.2\times10^{-6}$ \cite{mar}.

The registration of neutron emission in the real-time mode was performed by plastic scintillation detectors. For the detection of fast neutrons and X-rays POPOP doped polystyrene scintillators were used. The scintillators have an active area  of $15\times15$~cm$^2$ and the thickness of 5.5~cm. Signals were registered by 4-channel digital storage osciloscope with 1~GHz bandwidth. The intrinsic efficiency of the detector to fast neutrons, measured with the $^{252}$Cf source, was equal to $\eta_{\rm{int}}= 0.17$.

\section{Results}

\begin{figure*}
\resizebox{0.8\textwidth}{!}{
\includegraphics{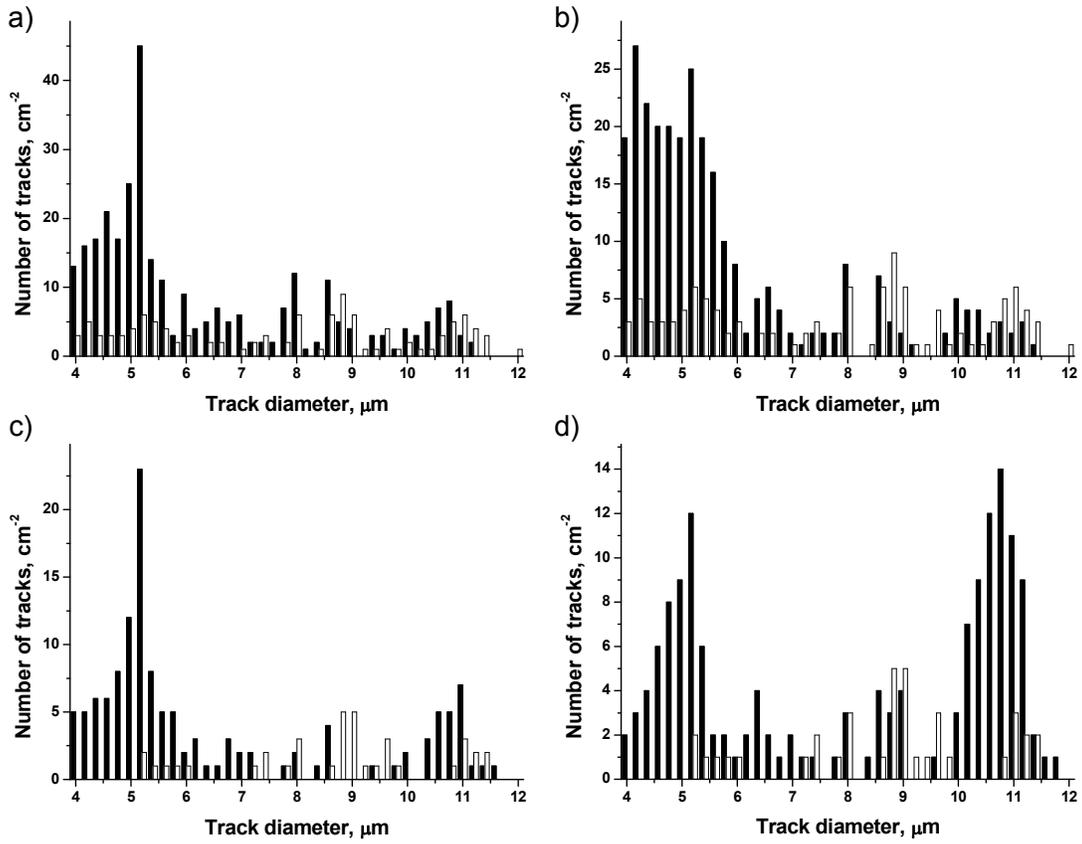}}
\caption{The total distribution of  track diameters: (a) and (b)
- detectors with radiators of 120~$\mu$m polyethylene, located in
the anode (a) and in the cathode (b); (c) - detectors with the
radiator of 20$\%$ solution Na$_2$B$_4$O$_7$ in glycerol, (d) -
detectors of the 20$\%$ solution of Na$_2$B$_4$O$_7$ in glycerol
with additional water moderator, placed near the anode. Dark
columns correspond to results of experiment, light - the relevant
background detectors, placed in 10~m from the discharge.}
\label{fig:2}
\end{figure*}

In a series of 180 shots the CR-39 track detectors, placed inside spherical anode and cathode, showed the excess of neutron flux over background. The track diameter distributions for the charged particles in the detectors are shown in Fig.~\ref{fig:2}. The round-shaped tracks with an angle of incidence close to the normal have been chosen. The CR-39 detector with PE radiator of 120~$\mu$m, placed inside the anode, revealed $\sim$5~time excess of tracks above the background within the range of recoil proton diameters of $4 - 6$~$\mu$m (Fig.~\ref{fig:2}a). This excess should correspond to the average fast neutron flux in the location of the detector $<\!\rm{n_{n1}}\!> \gtrsim 2\times10^5$~n$\cdot$cm$^{-2}$ per shot. A similar detector, located inside the cathode, also showed the excess above the background in the same range (Fig.~\ref{fig:2}b). The average flux of fast neutrons, estimated by the recoil protons, in the detector area was $<\!\rm{n_{n1}}\!> \gtrsim 3\times10^5$~n$\cdot$ cm$^{-2}$ per shot.

The detector inside the anode placed into a 20$\%$ solution of Na$_2$B$_4$O$_7$ in glycerol, has also showed the excess above the background (Fig.~\ref{fig:2}c) in $10 - 12$~$\mu$m range of diameters of $\alpha$-particle tracks (E$_\alpha < 2$~MeV). The average thermal neutron flux in the detector area, estimated by the reaction (\ref{eq1}), was $<\!\rm{n_{n2}}\!> \gtrsim 6\times10^5$~n$\cdot$cm$^{-2}$ per shot. Significant excess above the background in the $4 - 6$~$\mu$m range of the diameters of recoil proton tracks was also observed. Comparison of data from the detectors 1 and 3 has shown that the thermal neutron flux in the far detector (see. Fig.~\ref{fig:2}d) is greater. Probably, the fast neutrons are primary in the discharge, and then they slow down in the water moderator up to thermal energies.

\begin{figure*}
\resizebox{0.8\textwidth}{!}{
\includegraphics{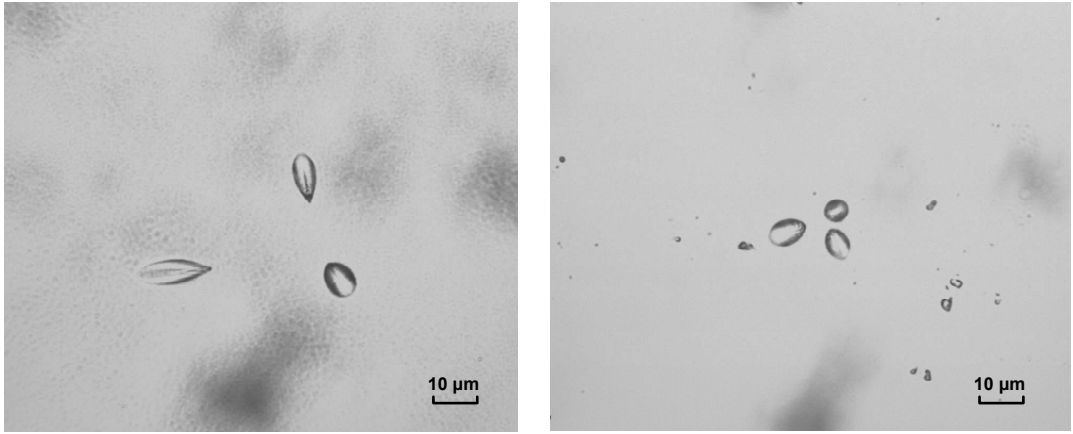}}
\caption{Typical photomicrographs of the events of $^{12}$C
nucleus desintegration into 3~$\alpha$-particles (the size of the
image $130\times100$~$\mu$m$^2$).}
\label{fig:3}
\end{figure*}

Very important result is connected with direct observation of desintegration of $^{12}$C nucleus into 3 $\alpha$-particles. Clear signature for such type events was observed on detectors located near discharge area. For example, at the detector 1 (2~cm$^2$), placed near the anode, there were observed 10 events of the desintegration of $^{12}$C nucleus into 3 $\alpha$-particles (Fig.~\ref{fig:3}). Therefore the number of 3$\alpha$-events was N$_{3\alpha} = 5$~cm$^{-2}$ at the background N$_{\rm{bg}3\alpha} = 0.2$~cm$^{-2}$, i.e. the confidence level of the effect of $^{12}$C nucleus desintegration is more than 10~$\sigma$. The average neutron flux of the energy E$_{\rm{n}} > 10$~MeV in the detector area, estimated by the reaction of (\ref{eq2}), was $<\!\rm{n_{n3}}\!> \gtrsim 1.4\times10^5$~n$\cdot$cm$^{-2}$ per shot. On the different detectors, located near the discharge, N$_{3\alpha} = 1 - 4$~cm$^{-2}$.

\begin{figure*}
\resizebox{0.6\textwidth}{!}{
\includegraphics{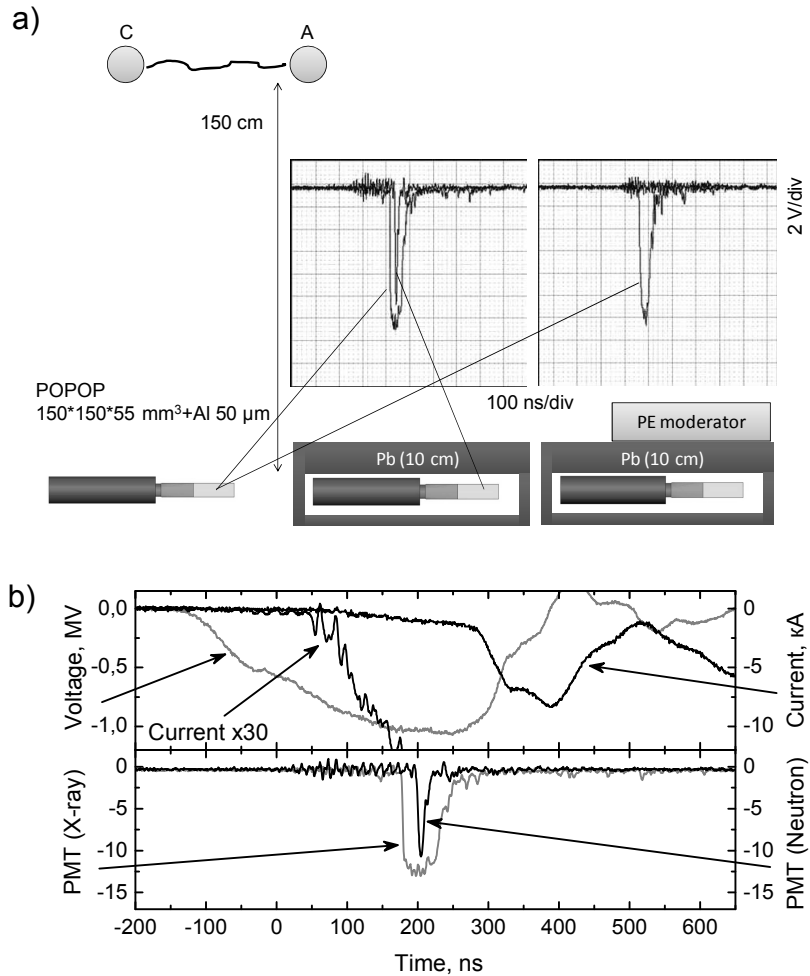}}
\caption{Scheme of real-time observation: (a) neutron signals, (b)
oscillograms of voltage, current, X-rays and neutrons.}
\label{fig:4}
\end{figure*}

The observation of neutron signals with plastic scintillation detectors is shown in Fig.~\ref{fig:4}a. The detectors are located at a distance of $\sim150$~cm from the discharge zone. One of detectors was shielded by 50~$\mu$m Al foil and could detect the pulses from the X-rays (E$_\gamma >$~10 keV) and neutrons (E$_n >$ 1~MeV). The other one had an additional 10~cm thick Pb shield. During a discharge both detectors indicated the time structure of signals. The start signal (presumably, from fast neutrons) in the second detector with the protection of 10~cm of Pb was delayed relative to the first signal (X-rays) by 35~ns. Using  time-of-flight estimation it was obtained the energy of neutron E$_{\rm{n}} \sim 10$~MeV. Taking into account the geometric factor ($8\times10^{-4}$) and detection efficiency, fast fraction of neutron flux was estimated  as $\sim (3-5) \times 10^5$  into 4$\pi$~sr per shot.

An additional neutron moderator (20~cm of polyethylene) leads to the absence of a signal from fast neutrons at the second detector. This is in agreement with the assumption of fast neutrons emission. As seen from Fig.~\ref{fig:4}b, the neutron bursts occur on the pre-pulse current prior to the formation of the discharge. At the moment of neutron generation the pre-pulse current amplitude is $\approx30$~times smaller then the amplitude of the main current  pulse.

\section{Conclusion}

Two independent methods (track detector CR-39 and the detector on the basis of plastic scintillators with PMT) have revealed that in the process of a high-voltage discharge in the air the neutrons are emitted in a wide energy range (from thermal up to the energies greater than 10~MeV) with the intensity of $\gtrsim10^6$~neutrons per shot into 4$\pi$ solid angle. The obtained data allows one to assume that during the discharge the fast neutrons are produced, and their generation occurs at the initial phase of the discharge and is correlated with the generation of X-ray radiation. To explain the mechanism of the observed emission of neutrons and clarify the location of their source one needs additional experiments.

\begin{acknowledgments}
The authors express their gratitude to S.M.~Zakharov, V.A.~Bogachenkov and E.I.~Saunin for assistance in conducting the experiments, A.V.~Gurevich, S.S.~Gershtein and G.A.~Mesyats for helpful discussions. The work was supported in part by the RFBR grant No.~13-08-01379.
\end{acknowledgments}

\end{document}